\documentclass[longauth,traditabstract,usenatbib]{aa} 
\usepackage{graphicx}

\usepackage{txfonts}

\usepackage{color}

\usepackage[nointegrals]{wasysym}

\usepackage{amssymb}

\usepackage[separate-uncertainty = true,multi-part-units=single]{siunitx}

\usepackage{natbib}
\bibpunct{(}{)}{;}{a}{}{,}

\PassOptionsToPackage{american}{babel}
\usepackage{babel}

\PassOptionsToPackage{authoryear}{natbib}
 \usepackage{natbib}

\renewcommand{\deg}{^\circ}

\usepackage[normalem]{ulem}
\newcommand{\Correction}[2]{\textcolor{red}{\sout{#1}} \textcolor{blue}{#2}}

\newcommand{\Revision}[1]{\textbf{#1}}

\usepackage{scrextend}

\begin{document} 

\title{Observations of Sagittarius A* during the pericenter passage of the G2 object with MAGIC}


%
\author{
M.~L.~Ahnen\inst{1} \and
S.~Ansoldi\inst{2} \and
L.~A.~Antonelli\inst{3} \and
P.~Antoranz\inst{4} \and
C.~Arcaro\inst{5} \and
A.~Babic\inst{6} \and
B.~Banerjee\inst{7} \and
P.~Bangale\inst{8} \and
U.~Barres de Almeida\inst{8,}\inst{24} \and
J.~A.~Barrio\inst{9}
% ...
}
\institute { ETH Zurich, CH-8093 Zurich, Switzerland
\and Universit\`a di Udine, and INFN Trieste, I-33100 Udine, Italy
\and INAF National Institute for Astrophysics, I-00136 Rome, Italy
\and Universit\`a  di Siena, and INFN Pisa, I-53100 Siena, Italy
\and Universit\`a di Padova and INFN, I-35131 Padova, Italy
\and Croatian MAGIC Consortium, Rudjer Boskovic Institute, University of Rijeka, University of Split and University of Zagreb, Croatia
\and Saha Institute of Nuclear Physics, 1/AF Bidhannagar, Salt Lake, Sector-1, Kolkata 700064, India
\and Max-Planck-Institut f\"ur Physik, D-80805 M\"unchen, Germany
\and Universidad Complutense, E-28040 Madrid, Spain
\and Inst. de Astrof\'isica de Canarias, E-38200 La Laguna, Tenerife, Spain; Universidad de La Laguna, Dpto. Astrof\'isica, E-38206 La Laguna, Tenerife, Spain
\and University of \L\'od\'z, PL-90236 Lodz, Poland
\and Deutsches Elektronen-Synchrotron (DESY), D-15738 Zeuthen, Germany
\and Institut de Fisica d'Altes Energies (IFAE), The Barcelona Institute of Science and Technology, Campus UAB, 08193 Bellaterra (Barcelona), Spain
\and Universit\"at W\"urzburg, D-97074 W\"urzburg, Germany
\and Institute for Space Sciences (CSIC/IEEC), E-08193 Barcelona, Spain
\and Technische Universit\"at Dortmund, D-44221 Dortmund, Germany
\and Finnish MAGIC Consortium, Tuorla Observatory, University of Turku and Astronomy Division, University of Oulu, Finland
\and Unitat de F\'isica de les Radiacions, Departament de F\'isica, and CERES-IEEC, Universitat Aut\`onoma de Barcelona, E-08193 Bellaterra, Spain
\and Universitat de Barcelona, ICC, IEEC-UB, E-08028 Barcelona, Spain
\and Japanese MAGIC Consortium, ICRR, The University of Tokyo, Department of Physics and Hakubi Center, Kyoto University, Tokai University, The University of Tokushima, KEK, Japan
\and Inst. for Nucl. Research and Nucl. Energy, BG-1784 Sofia, Bulgaria
\and Universit\`a di Pisa, and INFN Pisa, I-56126 Pisa, Italy
\and ICREA and Institute for Space Sciences (CSIC/IEEC), E-08193 Barcelona, Spain
\and now at Centro Brasileiro de Pesquisas F\'isicas (CBPF/MCTI), R. Dr. Xavier Sigaud, 150 - Urca, Rio de Janeiro - RJ, 22290-180, Brazil
% ...
}

   \date{Received XX XX, 2016; accepted XX XX, 2016}


 
  \abstract
   {We present the results of a multi-year monitoring campaign of the Galactic Center (GC) with the MAGIC telescopes. These observations were primarily motivated by reports that a putative gas cloud (G2) would be passing in close proximity to the super-massive black hole (SMBH), associated with Sagittarius A*, located at the center of our galaxy. This event was expected to give astronomers a unique chance to study the effect of in-falling matter on the broad-band emission of a SMBH.}  
   {We search for potential flaring emission of very-high-energy (VHE; $\geq$100 GeV) gamma rays from the direction of the SMBH at the GC due to the passage of the G2 object. Using these data we also study the morphology of this complex region.}
   {We observed the GC region with the MAGIC Imaging Atmospheric Cherenkov Telescopes during the period 2012-2015, collecting 67 hours of good-quality data. In addition to a search for variability in the flux and spectral shape of the GC gamma-ray source, we use a point-source subtraction technique to remove the known gamma-ray emitters located around the GC in order to reveal the TeV morphology of the extended emission inside that region.}
   {No effect of the G2 object on the VHE gamma-ray emission from the GC was detected during the 4 year observation campaign. We confirm previous measurements of the VHE spectrum of Sagittarius A*, and do not detect any significant variability of the emission from the source. Furthermore, the known VHE gamma-ray emitter at the location of the supernova remnant G0.9+0.1 was detected, as well as the recently discovered VHE source close to the GG radio Arc.}
   {}

  
  \titlerunning{MAGIC Observations of the Galactic Center region}
  \maketitle

%

\section{Introduction}
\subsection{The Galactic Center Region}

The central region of our galaxy is very densely populated with a large variety of astrophysical objects, many of which may be sites of extreme particle acceleration and hence gamma-ray emission \citep{vanEldik201545, aharonian2006hess, hess_galactic_ridge}. Multi-wavelength observations of this region and their interpretation have always been challenging due to a combination of source confusion and absorption along the line of sight \citep{genzel2010galactic}. 

Regardless, the Galactic Center (GC) region has been observed by several astronomical instruments over the previous three decades. The most precise data, especially regarding angular resolution, are coming from observations in the near IR \citep[$\SI{4e-2}{arcsec}$ resolution, ][]{genzel2003near} and radio \citep[$\SI{5e-4}{arcsec}$ resolution, ][]{SizeSgrA*_Bower} using large scale instruments like the Very Large Telescope (VLT, near IR), the Very Large Array (VLA, radio) and the Very Long Baseline Array (VLBA, radio). In the X-ray regime, the Chandra and NuStar satellites offer excellent angular (Chandra: 0.5~arcsec, NuStar: 9.5~arcsec) and energy resolution for the study of the GC region \citep{baganoff2000chandra,weisskopf2000chandra, Mori_NuStarGC, Kistler_2Pulsars}. 

These observations have revealed several astrophysical sources in the GC region. Among those, the compact radio source Sagittarius A* \mbox{(Sgr~A*)} is of a particular interest, and is generally accepted to be associated with the $\SI{4e6} M_{\astrosun}$ black hole at the center of our galaxy. The apparent size of the event horizon of the SMBH is estimated to be about $\SI{e-5}{arcsec}$ \citep{fish20111}.

In the X-ray domain, Sgr~A* is an unexpectedly faint emitter \citep[L$_x$ $\approx$ 2$\times$10$^{33}$ ergs s$^{-1}$ in the 2-10 keV band,][]{2003ApJ...BaganoffFaint} that does however display sporadic X-ray flaring activity on timescales from minutes to hours \citep{2001NatureBaganoff}. \cite{Ponti2015_XRay} presented an analysis of 15 years of X-ray observations (from September 1999 until November 2014) of Sgr~A* taken with the XMM-Newton and Chandra observatories. Interestingly, this study found an increase by a factor of 2-3 in the X-ray flare luminosity of Sgr~A* between 2013 and 2014 (although with a significance of only 3.5~$\sigma$), along with an increase in the rate of bright and very-bright X-ray flares with a significance of 3.3~$\sigma$. It should be noted that the authors acknowledge that this increase in measured flaring activity may purely be a sampling issue due to an increase in the monitoring frequency of Sgr~A* during that period.

The GC region has been also extensively observed in the high energy (HE; $\geq${100~MeV}) gamma-ray regime with the EGRET~\citep{EGRET_GC_1998} and \textit{Fermi}~\citep{GCFermi_2015} instruments, and in the very-high-energy (VHE; $\geq${100~GeV}) regime with Imaging Atmospheric Cherenkov Telescopes (IACTs). The first strong hints for a detection of VHE gamma rays from the GC were reported by CANGAROO~II~\citep{enomoto2003universe}, and one year later by the Whipple collaboration~\citep{kosack2004tev}. The H.E.S.S. collaboration in the same year reported a highly significant ($\sim$10$\sigma$) detection of a source at the GC with spectral index $\alpha$ = 2.2, designated as HESS J1745-290~\citep{aharonian2004very}. MAGIC observations confirmed these results with a compatible flux and spectral index~\citep{albert2006observation}. 

Recently, new observational results from H.E.S.S. and VERITAS have been published \citep{hess_collaboration_acceleration_2016, 2016VERITASGC_Smith, veritas2014very, aharonian2009spectrum}. In particular, analysis from the continued H.E.S.S. observations of the region around the GC suggest that the Sgr~A* black hole is able to accelerate particles to PeV energies~\citep{hess_collaboration_acceleration_2016}. Observations by VERITAS of the Galactic Center ridge have revealed the presence of a new source (VER J1746-289) near the GC~radio~Arc~\citep{2016VERITASGC_Smith}. Previously, H.E.S.S., MAGIC and VERITAS have reported on a new source of VHE emission from the same region near the GC \citep{2015HESSicrc,GC_ICRC_MAGIC,smith_veritas_2015}, which we will also address here.

The source and mechanism responsible for the production of HE and VHE gamma radiation from the GC still remain an active topic of discussion. Sgr~A* and the pulsar wind nebula G~359.95-0.04~\citep{2006_Wang_G359, 2007Hinton} are the leading candidates in the region since~\citet{aharonian2009spectrum} and~\citet{acero2010localizing} claimed to be able to rule out the nearby supernova remnant Sagittarius A East as a main contributor to the TeV emission. Several models for the production of high-energy radiation from Sgr~A* itself have been presented, including leptonic~\citep{2012Kusonose, 2004GCBH_Plerion}, hadronic~\citep{2012Fatuzzo,2012Linden,2011Ballantyne,2011Chernyakova,Wang_InjectionBH,aharonian_tev_2005} and hybrid \citep{2013Guo} scenarios.


\subsection{The G2 Object}

\citet{gillessen2012gas} reported the VLT infrared detection of a gas cloud with an estimated mass on the order of 3 Earth masses ($\sim10^{-5} M_{\astrosun}$)
 on a highly eccentric orbit towards the central SMBH of our galaxy. Extrapolating the orbit led to a predicted pericenter passage in mid-2013 at a distance of about 3100 Schwarzschild radii ($R_g$) from the SMBH. After continuous measurements these numbers were updated to September 2013 and 2200 $R_g$~\citep{G2_2013}. \citet{G2_2013_passage} reported that part of the gas cloud was observed past the pericenter approach by early 2013 and that the whole process would probably extend over at least one whole year. Other observations of the G2 object have resulted in the suggestion that G2 may in fact be the product of a binary-star merger \citep{Witzel_G2_Binary, Meyer_G2_KeckStellar} or a young star with $m \lesssim 3 M_{\astrosun}$~\citep{zajacek_infrared-excess_2015} as opposed to a gas cloud.

Predictions concerning the fate of the object and its possible influence on the accretion rate of the SMBH at the GC were highly dependent on the assumed density and structure of G2 as well as the physical environment close to Sgr~A*. These predictions ranged from no observable effects to strong flaring activity of Sgr~A*~\citep{schartmann2012simulations,giannios2013s2}. \citet{bartos2013g2} suggested that G2 may also interact with stellar-mass black holes expected to exist in the vicinity of Sgr~A*. 

Despite all of the uncertainties concerning the nature of this object, order-of-magnitude estimations can be made in order to put limits on the potential effects of a SMBH accretion event on the observed gamma-ray flux of Sgr~A*. The maximal amount of energy that can be released in the process of accretion of an object with mass $m$ onto a black hole is between 6\% and 42\% of the object's rest-mass energy ($mc^2$), depending on the black hole angular momentum~\citep{1983ShapiroBook}. Assuming that G2, with a mass of $\sim10^{-5} M_{\astrosun}$, is accreted onto Sgr~A* over the duration of a year (i.e. 30 times higher than the baseline accretion rate estimated by~\cite{yusef-zadeh_compact_2015}) and estimating the total accretion-disk luminosity to be $\sim10^{-1}\dot{M}c^{2}$ \citep[Equation~14.5.3 from][]{1983ShapiroBook}, the power released would be of the order of $10^{40.5}-10^{41.5}$~ergs s$^{-1}$. If only a small fraction of this power is used to accelerate high-energy particles, the resulting photon flux may well be comparable to the observed gamma-ray luminosity of the GC ($\sim 10^{35}$~ergs s$^{-1}$ above 0.5~TeV). 

Even considering the uncertainties in the predicted emission across all wavelengths, the possibility of observing in-falling matter onto the central SMBH of the galaxy was regarded as an interesting scientific opportunity and triggered MAGIC monitoring of the GC over the period 2012-2015. 

Despite the fact that recent observations in the near-infrared ~\citep{2014Ghez_Atel} appear to show that the G2 object has passed by Sgr~A* largely unaffected, and that observations reported by \cite{Radiomm_Monitoring_Bower} in the radio, millimeter and submillimeter wavebands taken during the apparent periastron passage of G2 show that the flux density and spectrum of Sgr~A* has remained stable (i.e. with measured flux density increases of ~20\% - consistent with typical low-luminosity Active Galactic Nuclei variability levels), the observational dataset accumulated by MAGIC still warrants a variability search in the VHE flux of the GC on a multi-year time scale. 

In the following sections, we report on the results of this 4-year observational campaign, covering the time period of the closest encounter between Sgr~A* and G2.

\section{The MAGIC observation campaign}

\subsection{The MAGIC Telescopes}

The MAGIC (Major Atmospheric Gamma Imaging Cherenkov) telescopes are two 17~m diameter IACTs, located at an altitude of 2200~m a.s.l. at the Roque de los Muchachos Observatory on the Canary Island of La Palma, Spain (28$^\circ$N, 18$^\circ$W).

The telescopes are used to record flashes of Cherenkov light produced by Extensive Air Showers (EAS) initiated in the upper atmosphere by gamma-ray photons with energies $\gtrsim$50 GeV. Both telescopes are nominally operated together in a so-called stereoscopic mode, in which only events simultaneously triggering both telescopes are recorded and analyzed~\citep{Magic_performanceII}. For low zenith distance (Zd) observations and for \mbox{$E>220$~GeV}, the integral sensitivity of MAGIC is $(0.66\pm0.03)\%$ in units of the Crab Nebula flux (C.U.) for 50 hours of observations~\citep{Magic_performanceII}. 

\subsection{Observations}
\label{sect:observations}

The GC region has been observed between April 2012 and July 2015, with 67 hours of good-quality data collected. When observed from the MAGIC site, the GC culminates at a zenith distance of $\mathrm{Zd} = 58^\circ$ and the time frame for observing the GC with MAGIC at Zd$<70^\circ$ is from mid-February until the end of September. A breakdown of the observation time per year, along with the relevant Zd range is shown in Table~\ref{tab:obs}.

\begin{figure}
	\centering
	\includegraphics[width=\columnwidth]{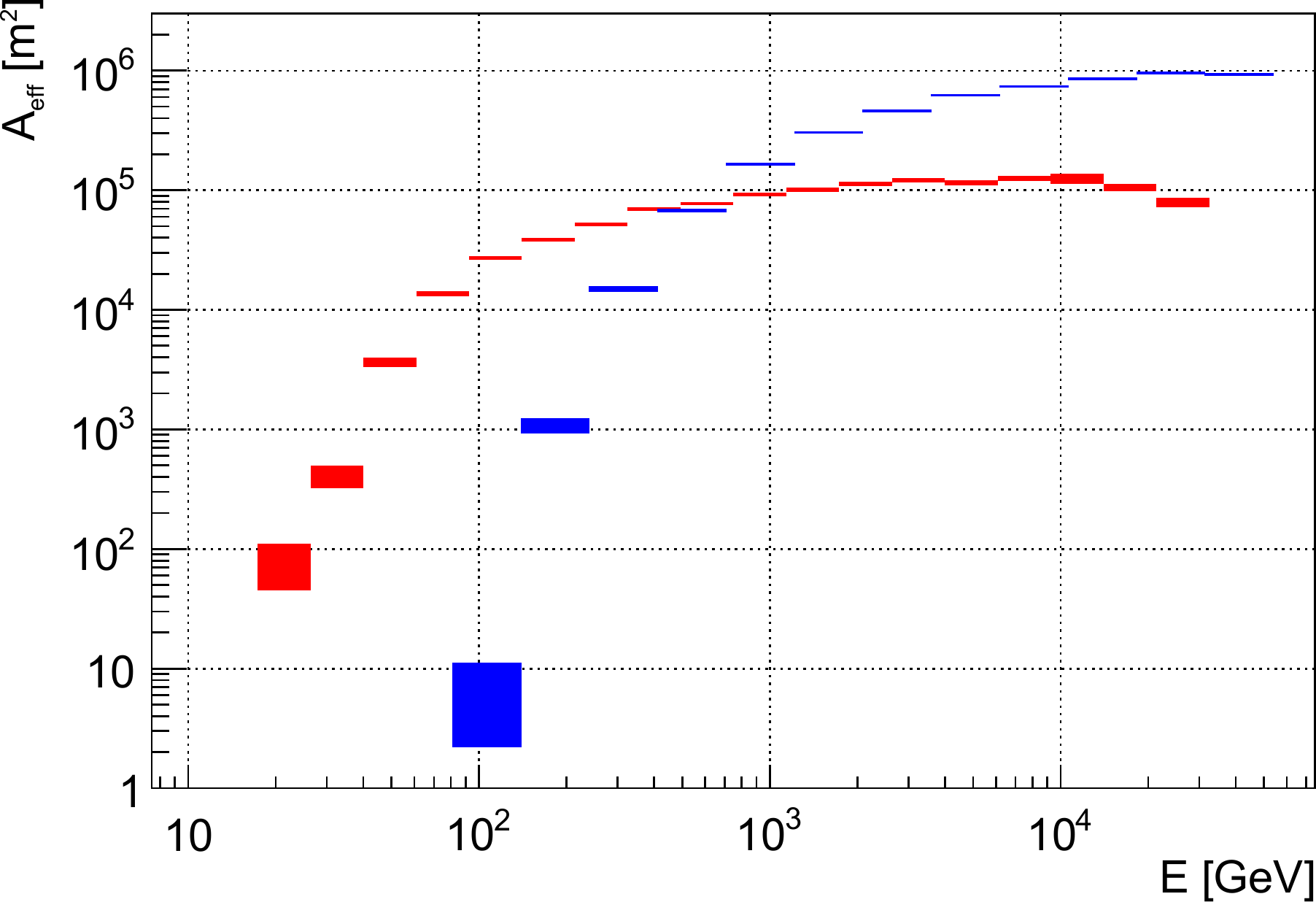}\\
	\caption{Effective collection area computed from Monte Carlo simulated events matching the zenith and azimuth distribution of the presented GC observation and after all cuts that were applied for computing the energy spectrum and light-curve (blue). The collection area for a typical low zenith angle ($5^\circ - 35^\circ$) observation is shown in red for comparison. One can clearly note the effect on the energy threshold and effective area due to the larger Zd.}
	\label{fig:col_area}
\end{figure}

Observing at such large zenith distances ($58^\circ$ to $70^\circ$ Zd) increases the energy threshold (defined as the peak in the distribution of detected gamma-ray events binned in energy, estimated using Monte Carlo simulations) of MAGIC to a range between $\sim$360~GeV and $\sim$1.2~TeV \citep[in general it varies with the zenith distance as $E_\mathrm{th,MAGIC} \sim \cos^{-2.3}{(Zd)}$, ][]{aleksic_major_2016}, but at the same time it also increases the effective collection area for gamma rays by nearly one order of magnitude.
\mbox{Figure~\ref{fig:col_area}} shows the comparison of two collection areas (post analysis-cuts) versus energy obtained from Monte Carlo simulations; corresponding to the zenith distance distribution covered by the MAGIC GC observations, and to a typical low-Zd ($5^\circ - {35}^\circ$) observation. 

The observations of the GC have been conducted in the False-Source tracking mode (also known as "Wobble" mode, ~\cite{1994Fomin_wobble}), meaning that the telescopes were pointed to four different symmetric positions at a distance of $0.4^\circ$ from Sgr~A*. With this observation technique, the background can be estimated from regions with the same camera acceptance. 
\begin{table}
	\caption{Summary of MAGIC GC observations by year. The listed observational times correspond to data surviving quality-selection cuts, as described in Sect.~\ref{sect:analysis} } 
	\label{tab:obs}                                   
	\centering                                        
	\begin{tabular}{l | r r r r}
		\hline\hline                        
		year                     &  2012 & 2013 & 2014 & 2015 \\
		\hline
		Obs. time [h]            &  3.0 & 25.9 & 27.2 & 11.2 \\
		Zd range [deg]           & 59 -- 66 & 59 --70 & 59 --70 & 58 -- 70 \\
		\hline                                             
	\end{tabular}
\end{table}

\subsection{Data analysis}
\label{sect:analysis}

The data have been analyzed with the MAGIC standard analysis chain MARS~\citep[MAGIC Analysis and Reconstruction Software,][]{zanin2013mars}. This chain includes the quality selection of the accumulated observations. During this step, the data are cleaned by removing events detected during periods of bad weather and/or during known temporary hardware issues. This basic data selection is performed based on several measured quantities, such as the mean photomultiplier currents, the event trigger rate, a measure of the amount of clouds in the field of view ~\citep[based on measurements with an infra-red pyrometer and the LIDAR system,][]{gaug_atmospheric_2014,fruck_novel_2014} and the number of stars detected by the MAGIC star-guider cameras during the observations.

Due to the nature of the large-Zd observations of GC (i.e. longer light path through the atmosphere), there is a larger impact on the quality of the recorded data due to a corresponding increase in the scattering of star light and decrease in the optical transmission of the atmosphere. To minimize these effects, strong quality cuts have been applied to the data. We have excluded periods of data taking when the photomultiplier currents were above twice the typical dark-night levels and also periods of data taking when the star-count reported by the star guider dropped below 70$\%$ of the median value. A cut on the data acquisition rate (dominated by the background cosmic-ray events) at $\pm 30\%$ of the typical value was applied as well, so that any data-taking periods when the event rates fluctuated substantially from the calculated mean rate were discarded. 

The remaining events were cleaned to remove the contributions of the night sky background and electronic noise. After that, the resulting shower images were parametrized in terms of the so-called Hillas~\citep{hillas_cerenkov_1985} and stereo parameters (\textit{disp}, shower height). Based on the MC simulated gamma rays and real background events, recorded in a sky region free of gamma-ray sources, the Random Forest technique~\citep{2012MAGICPerformance, Magic_RF} was used for event classification in order to substantially reduce the contribution of hadronic air-showers. 

Finally, an integration radius of $0.1^{\circ}$ around the coordinates of Sgr~A* (RA=17:45:40, Dec=-29:00:28) was used for extraction of the gamma-ray excess, which was later used to produce the energy spectrum and the light-curve of the source\footnote{Using this aperture photometry method, part of the extended emission from the GC ridge \citep{hess_galactic_ridge} may contribute at some degree to the measured flux.}. An aperture of fixed size has been used in order to minimize the effect of an energy dependent PSF in the context of an extended component in the source. The aperture size value of $0.1^{\circ}$ was also chosen to ensure that the results of this analysis can be more easily compared with previous measurements \citep[e.g.][]{aharonian2009spectrum,hess_collaboration_acceleration_2016}. The background rate within the $0.1^{\circ}$ integration radius has been calculated from a smoothed and modeled background estimation.

Shower images with size $< 50$~ph.e. (photo-electrons) were discarded from the analysis in order to remove poorly reconstructed events. For the morphological study of the GC region, an \emph{a-priori} cut on shower image size (200 ph.e. per telescope) was utilized to select only higher energy events ($E\gtrsim1$~TeV). This ensures that only well reconstructed events contribute to the sky maps shown in Figure \ref{fig:GC_skymap}, giving us a better signal-to-noise ratio and angular resolution. 

The background in the skymap is mostly caused by diffuse hadronic and electron events, and is estimated using the so-called Blind Map technique. This technique compares event rates for each bin in camera coordinates for different Wobble pointing positions and calculates the background model as the median for each of the pixels. In this way, regions that are affected by an increased number of counts due to a source contribution are automatically avoided. The merit of this method is that it does not rely on an a-priori knowledge of the source location (and extension) in the field of view. The caveat of this methodology is that sources with an extension larger than the distance between the Wobble positions (0.4~deg in radius, for these observations) would start contributing to the background model. At the same time this technique is suitable for point sources or moderately extended sources, regardless of their position in the FoV.

We have estimated the systematic uncertainty of our measurements based on~\citet{aleksic_major_2016}; yielding less than a 15$\%$ systematic uncertainty on the energy scale, 11$\%$-18$\%$ for the flux normalization and $\lesssim 0.02$~deg for the pointing accuracy. These numbers were determined at low and medium zenith distances and may therefore be underestimations for the data presented here due to the large average zenith distance of the observations. A~separate paper concerning this topic is currently in preparation. In the spectral energy distribution (SED) plots we show the effect of the systematic uncertainties by drawing four gray crossed arrows for different energy regimes. The vertical arrow indicates the systematic uncertainty on the flux scale. The systematic bias on the energy scale also leads to an error in the calculation of the SED, which depends on the shape of the collection area energy dependence. The resulting effect is depicted with the inclined arrows representing the influence of the systematic uncertainty of the energy reconstruction.


\section{Results}

After applying the quality cuts described in Section~\ref{sect:analysis}, the remaining \SI{67}{h} dataset yields a clear gamma-ray excess of events with $E>1$~TeV at the location of Sgr~A* with a significance of 27 standard deviations \citep[using formula 17 of][]{li_analysis_1983}.

\subsection{Gamma-ray emission spectrum}
\label{sect::spectrum}

\begin{figure}
	\centering
	\includegraphics[width=\columnwidth]{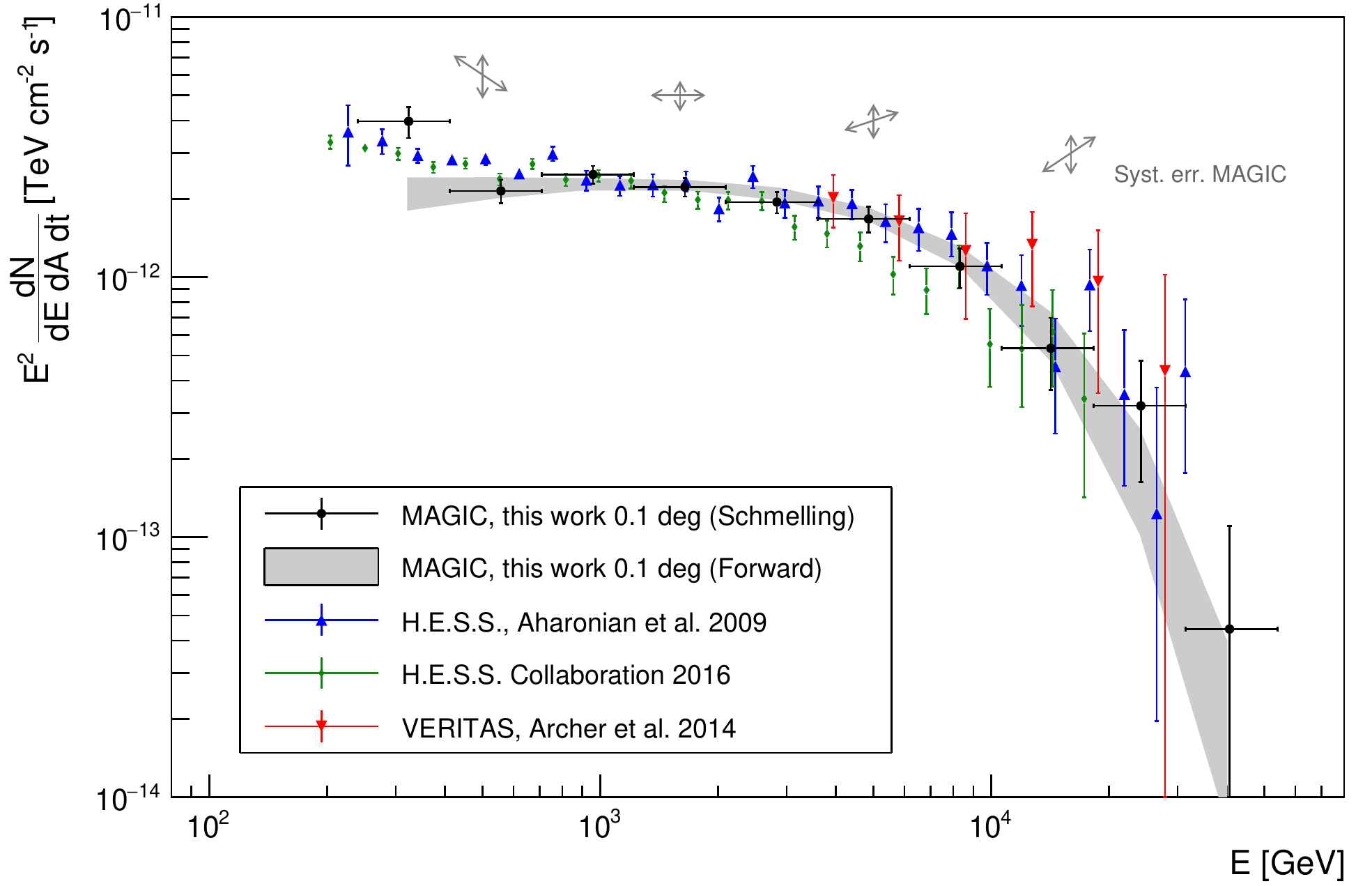}
	\caption{SED of the Galactic Center gamma-ray source measured by MAGIC between 2012 and 2015, unfolded with the method of~\citet{schmelling_method_1994} (black data points), and the forward folding fit result, assuming a power-law with exponential cut-off (shaded area, see Sect.~\ref{sect::spectrum} for details). Previous measurements by H.E.S.S and VERITAS are also shown for comparison. The gray arrows indicate the estimated systematic uncertainty of our measurements for different energy ranges (see Sect.~\ref{sect:analysis}), considering also the slope of the effective collection area vs. energy.}
	\label{fig:SEDs}
\end{figure}
The SED of Sgr~A* in the energy range $\SI{300}{GeV}-\SI{40}{TeV}$, unfolded with the method described in~\citet{schmelling_method_1994}, is shown in Figure~\ref{fig:SEDs}. The spectral shape has been found to be well described by a power-law with an exponential cut-off,
\begin{equation}
	\frac{\mathrm{d} F}{\mathrm{d} E} = f_0 \left(\frac{E}{E_0}\right)^{-\alpha}\, \exp\left({-\frac{E}{E_{cut}}}\right)
\end{equation}
The fit parameters of this model were obtained from the forward-folding fit to the measurements, which starts with the assumed spectrum and propagates it to detector counts using the response functions of the telescope. The latter included the MAGIC energy-migration matrix obtained from Monte Carlo simulations. This resulted in a fit with $\chi^2/NDF = 9.1/11$ (p-value is 0.61, $NDF$ stands for the number of degrees of freedom), and the following parameters: $f_0 = ({7.26 \pm 0.89})\times\SI{e-13}\\\SI{}{cm^{-2}s^{-1}TeV^{-1}}$, $\alpha = {1.85 \pm 0.13}$, $E_{cut} = ({7.57 \pm 2.29})\,\SI{}{TeV}$. The fit is normalized at $E_0 = \SI{2}{TeV}$. 

\begin{figure*}
	\centering
	\includegraphics[width=\textwidth]{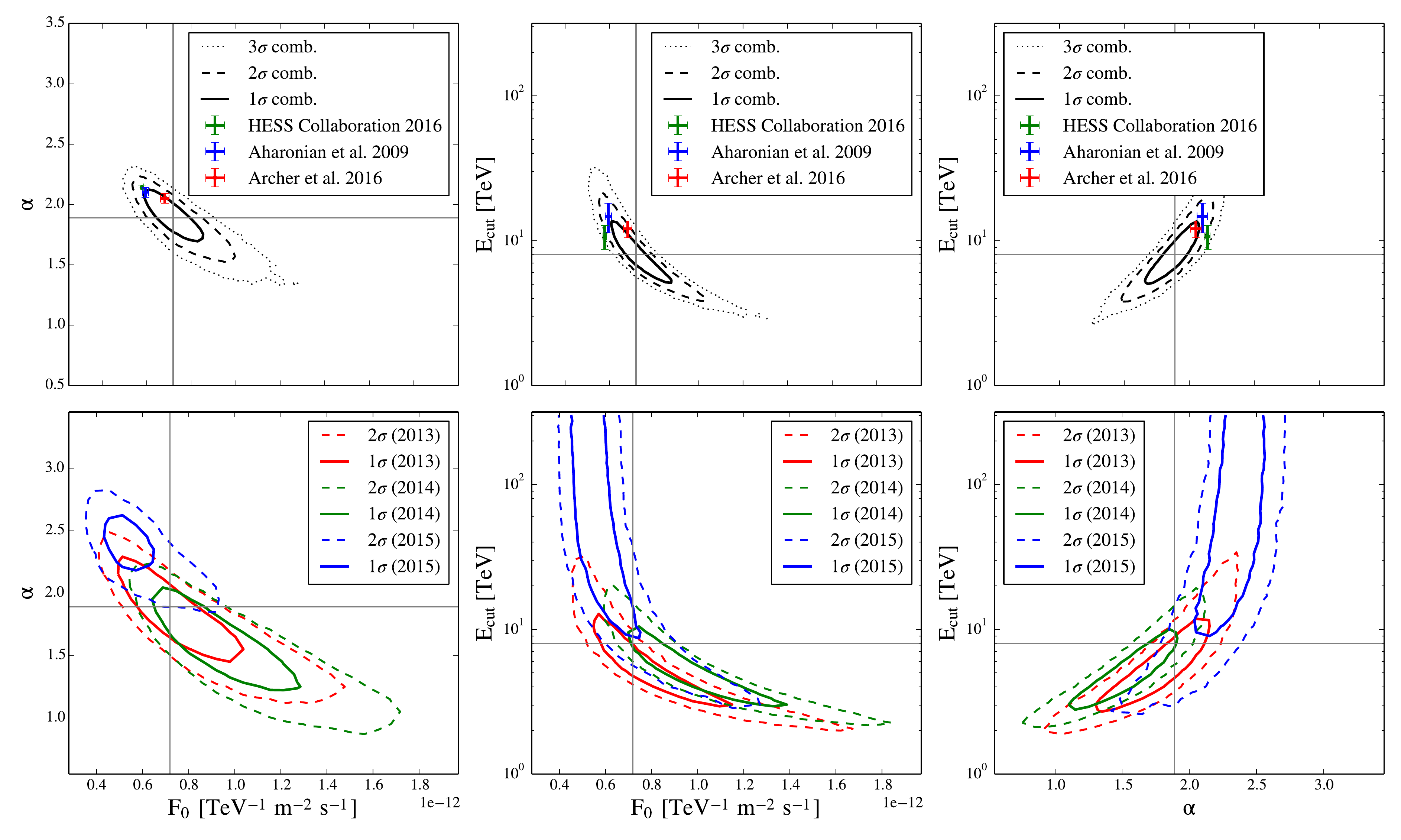}
	\caption{\textit{Top}: 2D projections of the probability distribution in the parameter space of the SED fit for the combined 2012-2015 data set. The SED was fit using a power-law with exponential cut-off model (see Sect.~\ref{sect::spectrum}). The gray cross-hair marks the best fit values found in this work. The red and blue data points mark the best fit values and uncertainties as measured by H.E.S.S.~\citep{aharonian2009spectrum} and H.E.S.S. with VERITAS combined~\citep{2016VERITASGC_Smith}. \textit{Bottom}: the same for the 2013 (red), 2014 (green) and 2015 (blue) seasons separately. The contours correspond to 1, 2 and 3 (only for the full data set, top) $\sigma$ confidence levels.}
	\label{fig:SED_parameter_crosscorr}
\end{figure*}

The above uncertainties should be treated with caution when used separately, as the fit parameters are significantly correlated between each other. To estimate this correlation we used a Markov chain Monte Carlo (MCMC) approach~\citep[\textit{emcee} algorithm:][]{goodman_ensemble_2010, foreman-mackey_2013} to sample the relevant parameter space and compute confidence contours. The sampling function used in the MCMC method was based on the Poissonian distribution, but was constructed to represent the likelihood of measuring a certain number of counts in the source region, given the model parameters and number of background events. The sampling also included the uncertainty on the MAGIC collection area, estimated from a dedicated Monte Carlo simulation which is part of the MAGIC standard analysis chain. The \textit{emcee} algorithm samples the parameter space with a large number of ``walkers'', reproducing the posterior probability density function (PDF) for $f_0$, $E_{cut}$ and $\alpha$ given the analyzed data sample. For simplicity in presentation, we have projected these distributions onto two-dimensional planes ``$f_0-\alpha$'', ``$f_0-E_{cut}$'' and ``$\alpha-E_{cut}$'', integrating over the third, remaining parameter. The resulting containment contours, corresponding to 1, 2 and 3$\sigma$ confidence levels, are shown in the top panel of Fig.~\ref{fig:SED_parameter_crosscorr}.

The obtained results are compatible with previous measurements by the H.E.S.S.~\citep{aharonian2009spectrum} and \mbox{VERITAS}~\citep{veritas2014very} experiments at $\sim 1 \sigma$ confidence level. At the same time we note that the updated H.E.S.S. spectrum of the source~\citep{hess_collaboration_acceleration_2016} deviates from our measurements by $\gtrsim 2 \sigma$.

\subsection{Search for variability}
\begin{figure*}
	\centering
	\includegraphics[width=0.8\textwidth]{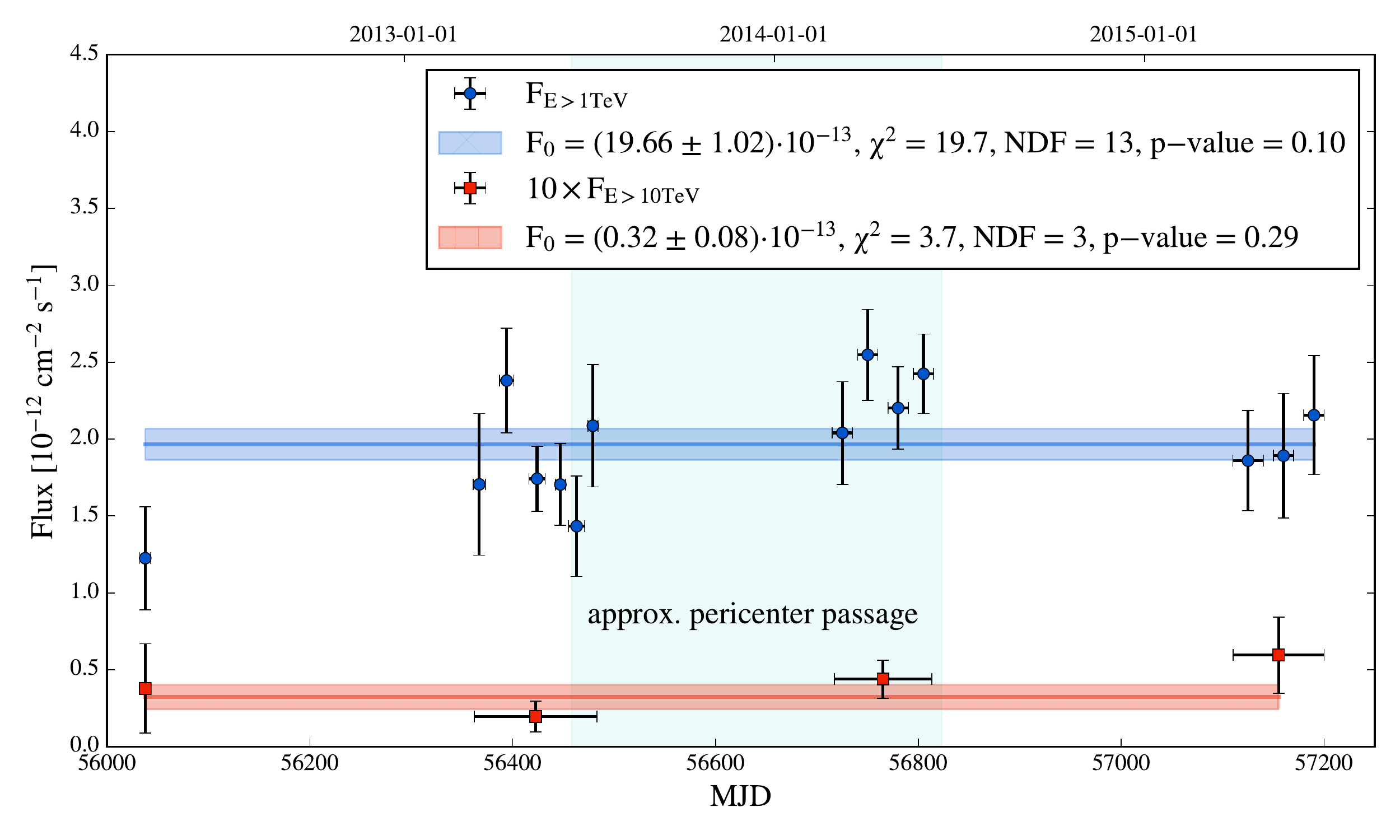}
	\caption{Light curves of the integral gamma-ray flux from the Galactic Center for $E > \SI{1}{TeV}$ and $E > \SI{10}{TeV}$. The red and blue lines represent the best fits to the constant flux, the corresponding shaded regions represent the $1\sigma$ confidence intervals. For $E > \SI{1}{TeV}$ the bins span over 10-30~days, whereas the yearly binning has been chosen for $E > \SI{10}{TeV}$ because of the low event count rate. The flux values of the latter have been multiplied by 10 for better visibility in the plot. Detailed information for each data point is given in Tables~\ref{tab:lc1} and~\ref{tab:lc10}.}
	\label{fig:LC_510}
\end{figure*}
We conducted a search for variability in the measured flux from Sgr~A* during the period of observations. Note that the predicted closest approach of the G2 object was to happen in 2013/14. Figure~\ref{fig:LC_510} shows the light curves of the integral flux \emph{F} for $E > 1$~TeV and $E > 10$~TeV, respectively. Detailed information about the individual measurements can be found in Tables~\ref{tab:lc1} and~\ref{tab:lc10}. In both cases the light curves are consistent with a constant flux assumption. For $E > \SI{1}{TeV}$, the fractional variability is less than $15\%$.
\begin{table*}
	\caption{Summary of the MAGIC flux measurements used in the light curve for $E > \SI{1}{TeV}$. } 
	\label{tab:lc1}                                   
	\centering                                        
	\small
	\begin{tabular}{l | c c c c c c c c c c c c c c  }
		\hline\hline
		Range [MJD-50000]  & 6032 -- 6043 & 6362 -- 6373 & 6387 -- 6401 & 6416 -- 6432 & 6442 -- 6452 & 6455 -- 6471 & 6474 -- 6483 \\
		\hline
		Observational time [h]        & 3.0 & 1.5 & 3.5 & 8.5 & 6.5 & 3.7 & 2.2 \\
		Detection significance [$\sigma$]& 4.9 & 4.1 & 7.5 & 9.2 & 6.6 & 5.2 & 5.6 \\
		Flux [$10^{-12}$s$^{-1}$cm$^{-2}$] & $1.23\pm0.34$ & $1.71\pm0.46$ & $2.38\pm0.34$ & $1.74\pm0.27$ & $1.70\pm0.27$ & $1.43\pm0.33$ & $2.09\pm0.40$ \\  
		\hline\hline 
		Range [MJD-50000] & 6717 -- 6735 & 6740 -- 6760 & 6770 -- 6790 & 6795 -- 6813 & 7110 -- 7140 & 7150 -- 7170 & 7180 -- 7200 \\
		\hline
		Observational time [h]        & 5.9 & 5.8 & 7.8 & 7.7 & 4.5 & 2.3 & 3.1 \\
		Detection significance [$\sigma$]& 5.9 & 9.3 & 8.8 & 10.1 & 6.1 & 5.4 & 6.4 \\
		Flux [$10^{-12}$s$^{-1}$cm$^{-2}$] & $2.04\pm0.33$ & $2.55\pm0.30$ & $2.20\pm0.27$ & $2.42\pm0.26$ & $1.86\pm0.33$ & $1.89\pm0.45$ & $2.15\pm0.39$ \\ 
		\hline
	\end{tabular}
\end{table*}
\begin{table*}
	\caption{Summary of the MAGIC flux measurements used in the light curve for $E > \SI{10}{TeV}$. }
	\label{tab:lc10}
	\centering  
	\small
	\begin{tabular}{l | c c c c}
		\hline\hline  
		Range [MJD-50000]      & 6032--6043 & 6363--6482 & 6719--6812 & 7110--7200 \\
		\hline
		Observational time [h]             & 3.0 & 25.9 & 27.2 & 11.2 \\
		Detection significance [$\sigma$]  & 1.0 & 1.8  & 4.3  & 2.7 \\
		Flux [$10^{-14}$s$^{-1}$cm$^{-2}$] & $3.5\pm3.4$ & $3.0\pm1.3$ & $5.2\pm1.5$ & $7.9\pm3.0$ \\
		\hline
	\end{tabular}
\end{table*}

Additionally, we also searched for signatures of any spectral variability of the source during the periods of observation. To achieve this for the seasons 2013, 2014 and 2015 we separately fitted the spectrum of Sgr~A* and compared the obtained parameters (the 2012 season was not fit due to the limited dataset). We used the MCMC approach described above to sample the parameter space for each season separately. The parameter cross-correlation diagrams for three years of MAGIC observations are shown in the lower panel of Figure~\ref{fig:SED_parameter_crosscorr}. They do not show any significant variation between the different observational seasons.

\subsection{Morphology of the emission}

\begin{figure*}
	\centering
	\includegraphics[width=0.5\textwidth]{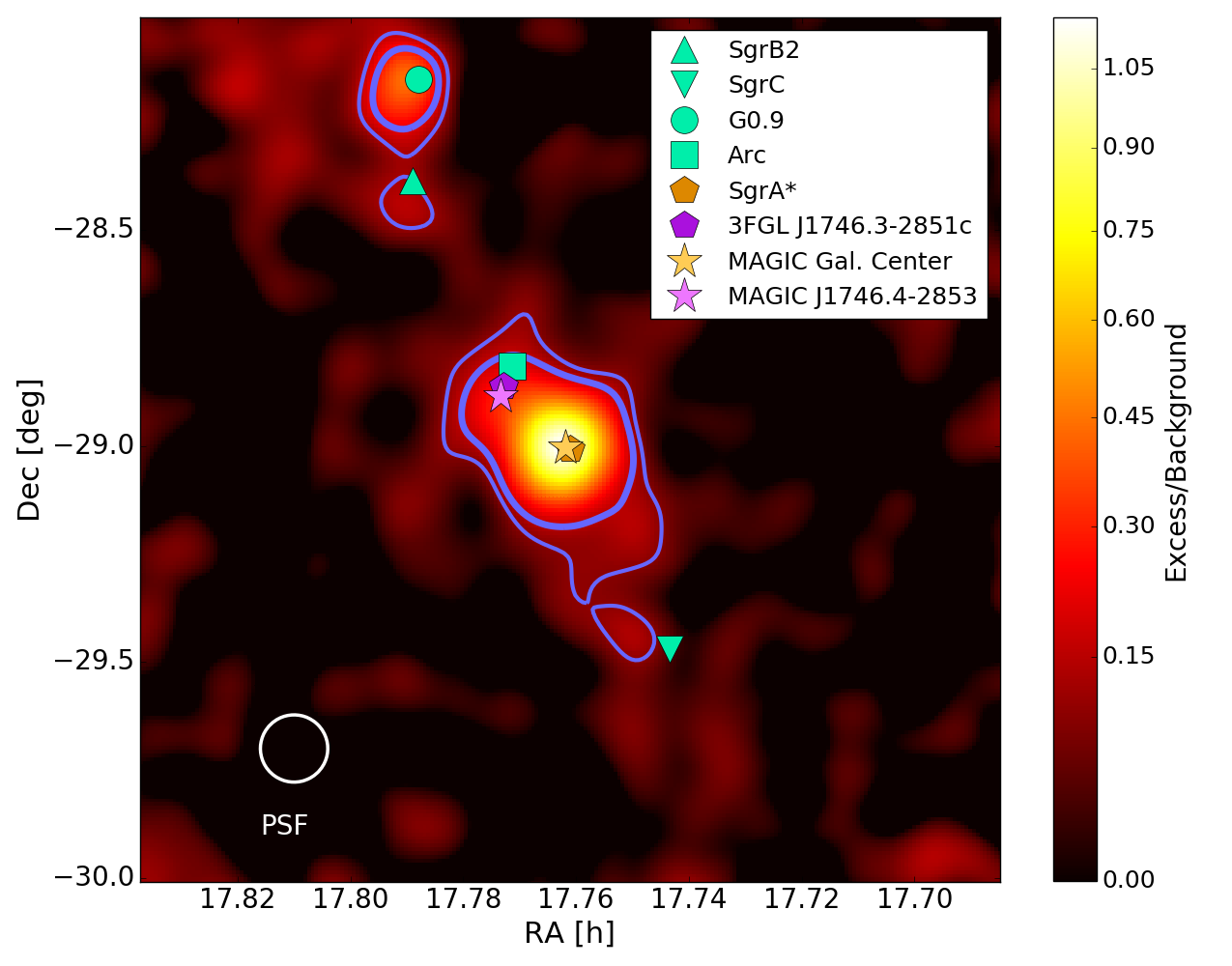}
	\includegraphics[width=0.5\textwidth]{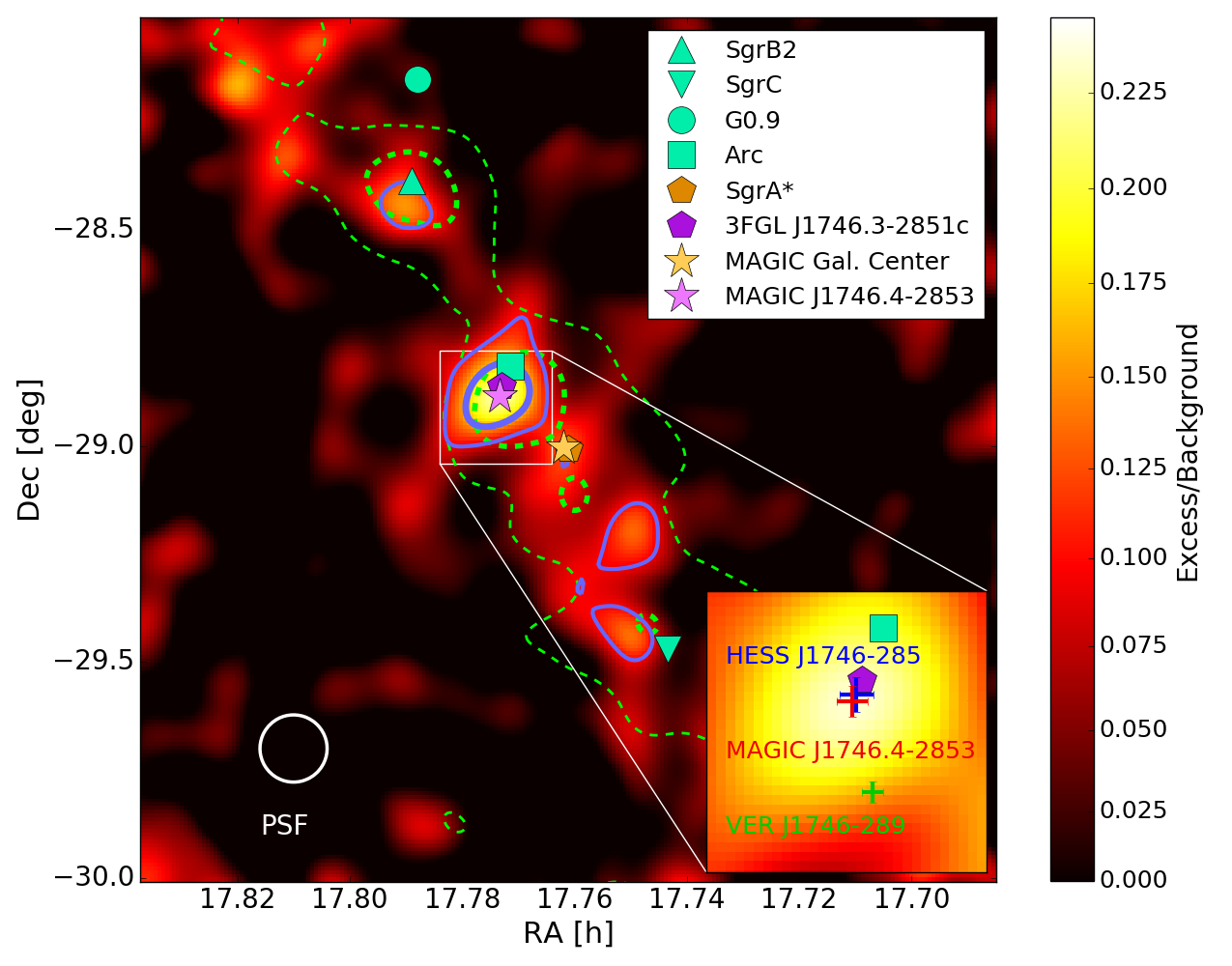}
	\caption[]{
	\textit{Left}: A sky map of the central $2~\mathrm{deg} \times 2$~deg field of view around the Galactic Center position, showing the relative signal to background count rate for $E \gtrsim \SI{1}{TeV}$.
	\textit{Right:} the same, with the Sgr~A* and G0.9+0.1 point sources subtracted from the map (see the details in the text). Both sky maps have been smeared with a Gaussian kernel ($\sigma_{ker}$ = 0.055${\deg}$) and are given in units of relative counts (number of excess events over the number of estimated background events using the Blind Map method). The blue contours show 3$\sigma$ (thin) and 5$\sigma$ (thick) local significance levels. Coordinates of the known radio structures are indicated with light green markers. The best-fit coordinates of the Galactic Center source and of the unidentified source (here referred to as MAGIC~J1746.4-2853) are indicated with stars. The coordinates of Sgr~A* (radio) and the \textit{Fermi} source 3FGL~J1746.3-2851c are indicated by pentagons. The MAGIC PSF is given as a $1 \sigma$ contour of a 2D~Gaussian smeared with the same kernel that was used for the sky map. For comparison, the H.E.S.S. contours at the event count levels of 320 and 360 are shown as dashed lines~\citep{hess_galactic_ridge}. Please note that the H.E.S.S contours correspond to the energy threshold of 380~GeV -- significantly different from that of MAGIC ($\sim 1$~TeV) for these skymaps. The inset in the lower right shows a zoom-in onto the Arc region with best fit coordinates of MAGIC~J1746.4-2853, VER~J1746-289, and HESS~J1746-285 shown as error crosses \citep{2016VERITASGC_Smith, 2015HESSicrc}, describing statistical errors from the fit only\footnotemark.}
	\label{fig:GC_skymap}
\end{figure*}
The region within one degree around the GC contains a collection of known gamma-ray sources. Apart from the point-like source component at the coordinates of Sgr~A*, MAGIC also detects emission from the known composite supernova remnant G0.9+0.1~\citep{collaboration_very_2005}, at the level of $9\sigma$ local significance. A sky map of the GC region as seen by MAGIC is provided in Figure~\ref{fig:GC_skymap} (left), which shows the relative count number (with respect to the remaining background) of the gamma-ray events with $E \gtrsim \SI{1}{TeV}$. The extended emission becomes more obvious if the gamma-ray flux contributions from Sgr~A* and G0.9+0.1 are removed from the image.
For this we subtract a sum of two symmetric 2D Gaussians -- the PSF model, which provides a reasonable description of the MAGIC point spread function~\citep{Magic_performanceII}, from the fitted coordinates of Sgr~A* and G0.9+0.1. The shape parameters (kernel of the first 2D Gaussian $\sigma_1 = 0.048 \pm 0.007^\circ$, kernel of the second 2D Gaussian $\sigma_2 = 0.092 \pm 0.015^\circ$, normalization ratio in terms of the second component $N_2/N_{tot} = \SI{0.51 \pm 0.18}{}$) of the PSF model have been determined with a $\chi^2$ fit to Crab Nebula data recorded at a similar Zd.
The result of the subtraction is shown in the right panel of Figure~\ref{fig:GC_skymap}. The residual sky map shows the extended emission from the region along the Galactic plane, similar in shape with the earlier findings~\citep[shown as dashed green contours][]{hess_galactic_ridge}, though detected here at higher energies.

In addition, a source of significant VHE gamma-ray emission located close to the Galactic Center Radio Arc (GCA)~\citep{1997Tsuboi}, 0.2$^\circ$ to the east of Sgr~A*, has been detected at a $7.2\sigma$ local ($6.4\sigma$ post-trials) significance level. The significance was evaluated through a test statistic based on the background emission that corresponds to the position of the source in terms of camera coordinates but is measured in the Off-source region. The VHE excess is consistent with a point-source at the coordinates RA~17:46:25, Dec~-28:52:55 with an error circle of $0.03$~deg, determined by fitting a single 2D~Gaussian shape. Throughout this paper we refer to this source as MAGIC~J1746.4-2853. We note that it is positionally consistent with the VHE excess
VER~J1746-289, recently reported by \cite{2016VERITASGC_Smith}, and previously presented by the MAGIC \citep{GC_ICRC_MAGIC}, HESS \citep{2015HESSicrc} and VERITAS \citep{smith_veritas_2015} collaborations.
Also the EGRET source 3EG~J1746-2851 and the \textit{Fermi} source 3FGL~J1746.3-2851c are in spatial coincidence with the VHE source. 

The inset in the right panel of Figure~\ref{fig:GC_skymap} shows the best fit coordinates of MAGIC~J1746.4-2853, VER~J1746-289, and HESS~J1746-285, with error bars containing the 68$\%$ C.L. (90$\%$ in case of HESS~J1746-285) statistical-fit uncertainty only. The errors for the other two sources have been taken from~\cite{2016VERITASGC_Smith} and ~\cite{2015HESSicrc} but rotated from the Galactic to Equatorial coordinate frame. The systematic pointing error of MAGIC is estimated to be $\lesssim 0.02$~deg, while \citet{2016VERITASGC_Smith} state a systematic pointing error of 0.013~deg in both Galactic latitude and longitude. 

The origin of this new source is unclear, though several possible associations with known objects can be speculated upon. One possible candidate is the giant molecular cloud (GMC) G0.11-0.11, located very close to the southern half of the GCA, and between the GCA and Sgr~A*. Gamma-ray emission could either originate from electrons accelerated in the interaction of G0.11-0.11 with the GCA \citep[such a scenario was already discussed by][]{pohl_galactic_1997}, or from CR interactions inside the dense molecular material in the region. Those cosmic rays could either originate from past active episodes of Sgr~A*, several hundreds or thousands of years ago, or they could have been accelerated in shocks associated with the numerous supernova explosions that have been driving the expansion of the GMC~\citep{oka_molecular_2001}. The analysis of the X-ray data suggests, alternatively, a possible association with a pulsar wind nebula candidate found within the positional uncertainty of the source~\citep{2015HESSicrc}. 

\section{Discussion}

The primary motivation behind this observing campaign was to search for any flaring emission that may occur due to the passage of the G2 object near to the SMBH at the center of the Milky Way galaxy. The proximity of the passage of the G2 object to the SMBH could have provided a unique opportunity to study the process of accretion of an Earth-mass body onto a black hole, as well as addressing several questions regarding particle-acceleration mechanisms near to a SMBH. However, the results of recent observations at other wavelengths suggest that the G2 object has not been disrupted by its proximity to the SMBH, therefore it is perhaps not surprising that no evidence for an enhancement in the VHE flux of Sgr~A* was found.  

Regardless, 10~years after the discovery of VHE emission from the region, the nature of the $\gamma$-ray source at the GC remains uncertain. The MAGIC observational campaign also aimed to help clarify this issue, by measuring both the overall spectral shape and variability of Sgr A* in the energy range above several hundreds of GeV.
\footnotetext{The statistical pointing errors of the best fit coordinates for VER~J1746-289 and HESS~J1746-285 in Figure \ref{fig:GC_skymap} have been obtained by transformation of the values, which are given in Galactic coordinates, into the Equatorial system by coordinate rotation. The errors on the position of HESS~J1746-285 correspond to 90$\%$ C.L.}

\begin{figure*}
	\centering
	\includegraphics[width=0.95\textwidth]{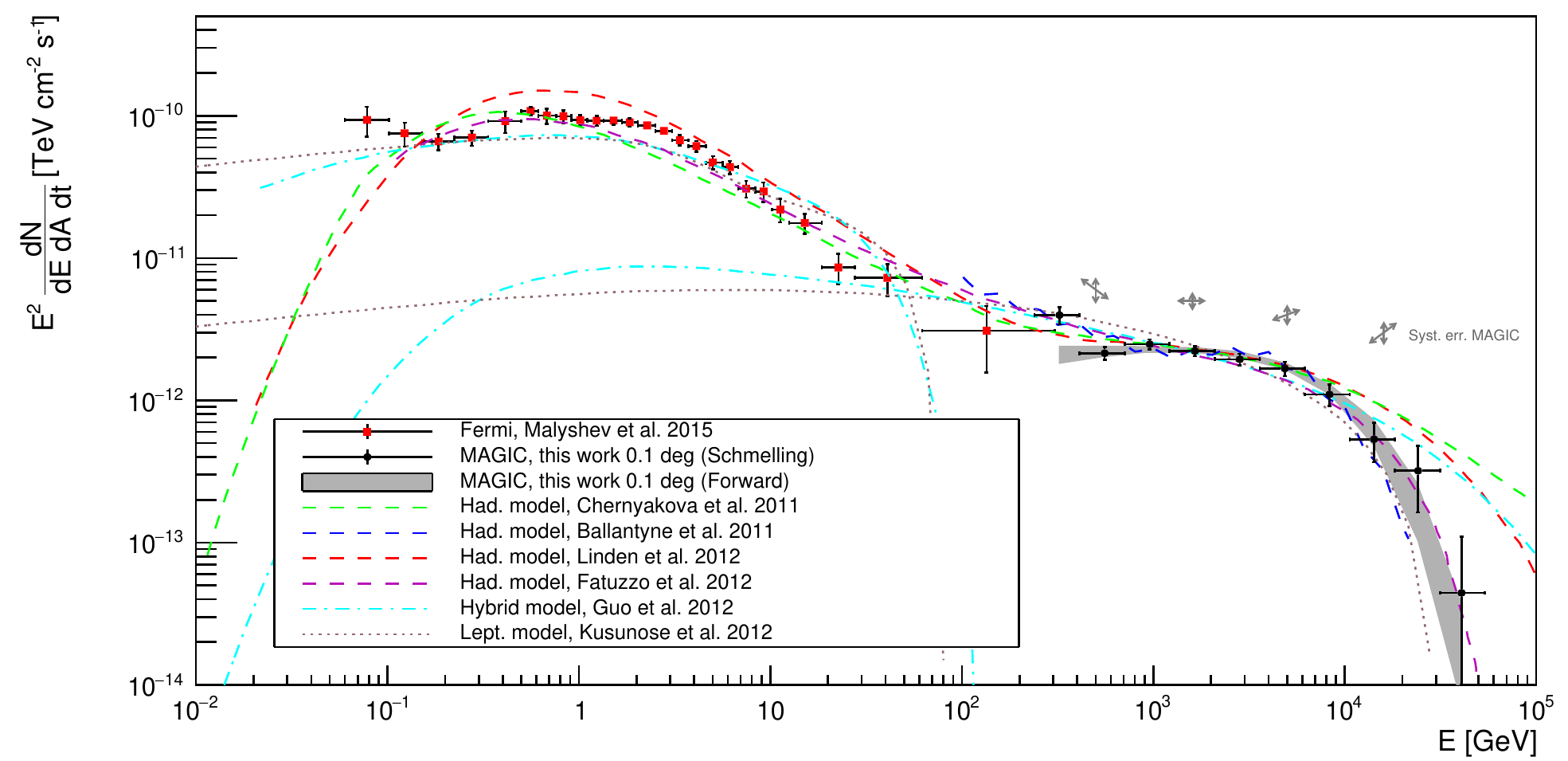}
	\caption{The GeV-TeV SED of SgrA*. The \textit{Fermi} data points in the $\sim 100\mathrm{MeV}-100\mathrm{GeV}$ band are from the most recent spectrum of~\citet{malyshev15}. Contemporary hadronic (dashed), leptonic (dotted) and hybrid-type (dash-dotted) models are shown for comparison.
	}
	\label{fig:SEDs_2}
\end{figure*}
The theoretical expectations for the spectral shape and flux variability significantly vary between the different assumed scenarios. Before the publication of the \textit{Fermi} spectrum on Sgr~A*~\citep{2011Chernyakova}, the models were built mainly around the TeV emission observed by H.E.S.S.~\citep{aharonian2004very,aharonian2009spectrum}. The MAGIC observations presented here confirm the previous measurement of the source SED and extend it up to $\sim 50$~TeV, providing a new test for both hadronic and leptonic type scenarios, proposed for explaining the observed VHE emission from Sgr~A*, as shown in Fig.~\ref{fig:SEDs_2}.

In most of the hadronic scenarios, the gamma rays are produced by $\pi^{0}$ decay from the interactions of cosmic rays (CRs), accelerated in the vicinity of the SMBH, with the dense environment close to the GC. In the model of~\citet{2011Ballantyne}, the measured TeV spectrum is obtained by switching on and off CR acceleration close to the SMBH at specific times in the past. The energy dependence of the diffusion coefficient is then responsible for the spectral shape. One implication of the~\citet{2011Ballantyne} model is that variability of the TeV spectrum ($\gtrsim$10 TeV) is expected on time scales of the order of 10 years, not only in the case that the accelerator stays quiet, but also if a new episode of CR acceleration occurs. According to the MAGIC results, there is no strong evidence for variable emission from Sgr~A* at these energies over the years 2012 -- 2015, as well as with respect to previous measurements.

After discovery of a point-source in the \textit{Fermi} data (1FGL~J1745.6-2900), which could be associated with the H.E.S.S. TeV source (HESS~J1745-290), \citet{2011Chernyakova} and later~\citet{2012Linden} proposed similar hadronic models, able to explain both the GeV and the TeV emission. These models use the injection spectrum resulting from CR acceleration close to the SMBH with a spectral index $\alpha \sim 2$ and an exponential cut-off at $\sim 100$~TeV. The variation of the spectral index of the gamma-ray emission along the spectrum is explained by the difference in the diffusion times for GeV and TeV cosmic rays. Both models are assuming a dense configuration of interstellar gas at distances from one to a few parsecs away from Sgr~A*.

In their hadronic model, \citet{2012Fatuzzo} include a simplified description of the particle acceleration in their numeric simulation of the diffusion of CRs through the turbulent magnetic fields expected around Sgr~A*. They assume a torus of dense material around the GC SMBH of about 2~pc in radius, embedded inside a wind zone of lower density, about 10~pc in diameter. Particles are accelerated throughout their diffusion history and eventually react with the ambient protons, either in the torus (generating HE emission) or in the wind zone (generating VHE emission). This model does not need time variability to explain the overall shape of the GeV-TeV spectrum.

Alternatively,~\citet{2012Kusonose} suggested a model where high-energy electrons are accelerated close to the central SMBH and interact via inverse-Compton scattering with soft photons, emitted by the dense population of stars and dust inside the central few parsecs of the GC. 
The electron populations would have to originate from different acceleration mechanisms or sources. A similar scenario is also suggested by~\citet{2007Hinton}.

A hybrid lepto-hadronic scenario was also recently suggested by~\citet{2013Guo}. In their model both electrons and protons are accelerated in the vicinity of the SMBH. The GeV part of the spectrum is attributed to the inverse-Compton scattering of relativistic electrons on the soft background photons, while the TeV emission is produced via the CRs colliding with the surrounding gas.

Hadronic scenarios have recently gained support through the measurement of gamma rays with energies up to over 40~TeV, which the authors \citep{hess_collaboration_acceleration_2016} interpreted as evidence for the presence of PeV protons in the region. Despite the temptation to link their presence to Sgr~A*, this is not straightforward due to the required energetics, exceeding the current bolometric luminosity of the source and the availability of alternative scenarios~\citep{hess_collaboration_acceleration_2016}. Regardless, such a connection is still considered very likely.

Considering the statistical and systematic errors on the Sgr~A* spectrum as measured by MAGIC, no single emission model can be unequivocally ruled out. See Figure~\ref{fig:SEDs_2} for an overview of contemporary modelling attempts presented alongside MAGIC and recent \textit{Fermi} data. The SED is in a reasonable agreement with the leptonic and hybrid type models, shown with the dotted~\citep{2012Kusonose} and dash-dotted~\citep{2013Guo} lines. Hadronic models seem to conflict with the lowest energy (60-100~MeV) \textit{Fermi} measurements ~\citep{malyshev15}.
However, to be able to distinguish between the various models, the study of flux variations over time, as predicted by most of the hadronic models, will have the highest separation power. So far the \textit{Fermi} observations in the GeV band have not yet revealed any significant variability~\citep{2011Chernyakova} and the MAGIC monitoring in the TeV band presented here, also measures a stable source flux. However it is still necessary to continue monitoring Sgr~A*, especially at the highest energies, where the most rapid variability on a timescale of the order of 10 years is predicted~\citep{2011Ballantyne}. For now, the absence of any detection of variability prevents the use of these measurements to disentangle the various emission models from each other. 

\section{Conclusions}

The GC region has been observed with the MAGIC telescopes between 2012 and 2015, collecting 67 hours of good-quality data. No effect of the G2 object on the VHE gamma-ray emission from the GC was detected during the 4 year observation campaign. The lack of variability from the direction of Sgr A*, as measured by MAGIC, makes it difficult to rule out single models describing particle acceleration and gamma-ray emission mechanisms at the source. These observations may still prove useful as an accurate measurement of the baseline emission from Sgr A* in the case of any possible flaring activity in the future.

Along with the variability study, the large exposure of 67 hours allowed us to derive a precise energy spectrum of Sgr~A*, which agrees with previous measurements within errors. Furthermore we were able to study the morphology of the GC region. As a result of this study, we confirm the detection in the VHE gamma-ray band of the supernova remnant G0.9+0.1, and report the detection with MAGIC of a VHE source of unknown nature in the region of the GC Radio Arc.

\begin{acknowledgements}
\newline
We would like to thank
the Instituto de Astrof\'{\i}sica de Canarias
for the excellent working conditions
at the Observatorio del Roque de los Muchachos in La Palma.
The financial support of the German BMBF and MPG,
the Italian INFN and INAF,
the Swiss National Fund SNF,
the he ERDF under the Spanish MINECO
(FPA2015-69818-P, FPA2012-36668, FPA2015-68278-P,
FPA2015-69210-C6-2-R, FPA2015-69210-C6-4-R,
FPA2015-69210-C6-6-R, AYA2013-47447-C3-1-P,
AYA2015-71042-P, ESP2015-71662-C2-2-P, CSD2009-00064),
and the Japanese JSPS and MEXT
is gratefully acknowledged.
This work was also supported
by the Spanish Centro de Excelencia ``Severo Ochoa''
SEV-2012-0234 and SEV-2015-0548,
and Unidad de Excelencia ``Mar\'{\i}a de Maeztu'' MDM-2014-0369,
by grant 268740 of the Academy of Finland,
by the Croatian Science Foundation (HrZZ) Project 09/176
and the University of Rijeka Project 13.12.1.3.02,
by the DFG Collaborative Research Centers SFB823/C4 and SFB876/C3,
and by the Polish MNiSzW grant 745/N-HESS-MAGIC/2010/0.
\end{acknowledgements}


\bibliographystyle{aa}
\bibliography{GCG2_References}

\begin{appendix}
\section{Energy migration and unfolding of the MAGIC spectrum}
\label{sect::appendix}

\begin{figure}
	\centering
	\includegraphics[width=\columnwidth]{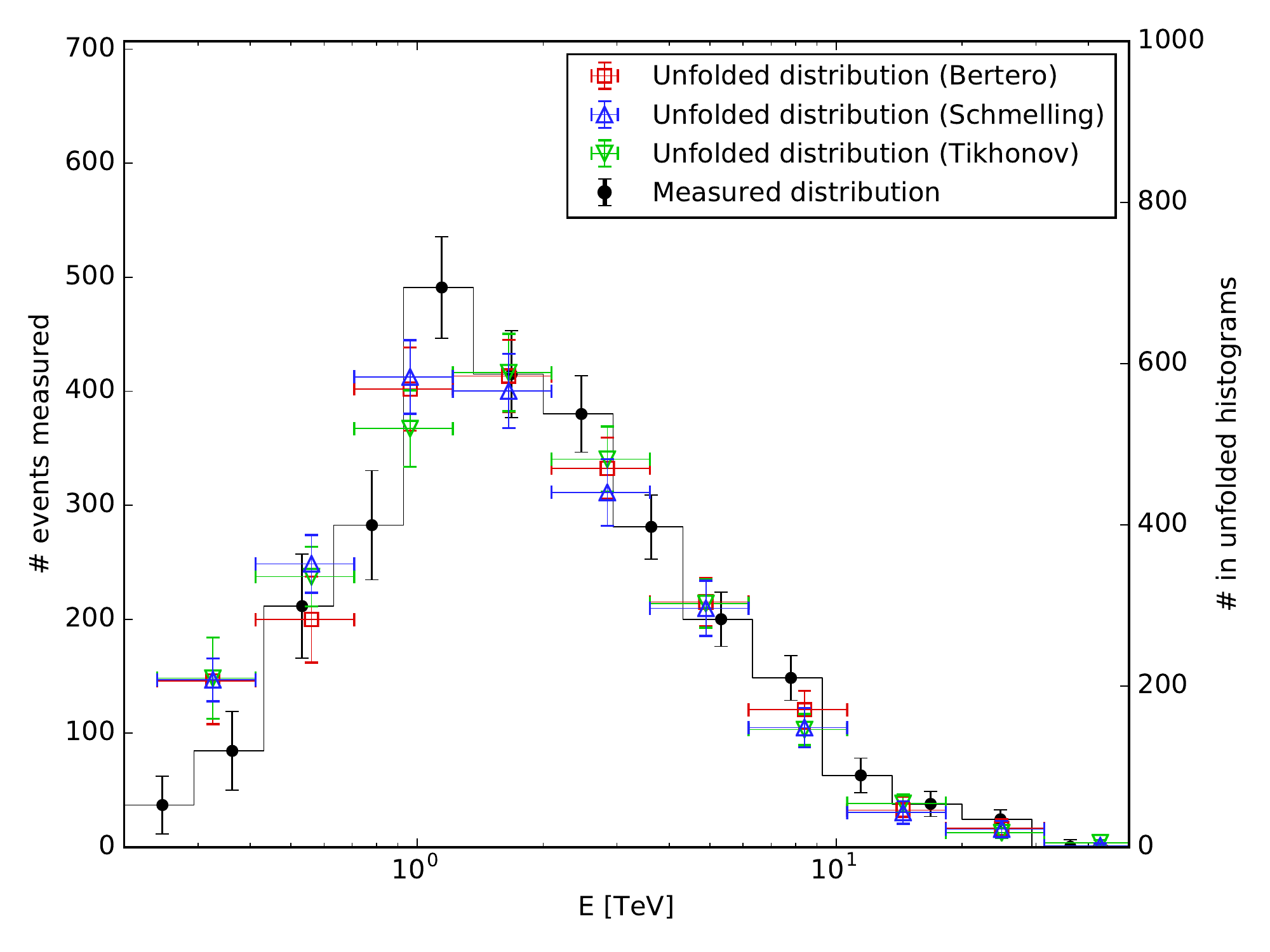}
	\caption{Energy distribution of the excess events in the SgrA* data sample. The measured event distribution in terms of estimated energy is shown in black and is accompanied with the corresponding unfolded distributions, displayed in ``true'' energy. The details of the procedure are given in Appendix~\ref{sect::appendix}.}
	\label{fig:EvtHistos}
\end{figure}

The reconstruction of the energy of the primary gamma ray that initiated an air-shower has limited accuracy. The finite energy resolution results in migration of events between the neighbouring energy bins, which may lead to significant spillovers from more to less populated energy bins. The measured event energy distribution for the MAGIC GC data set -- subject to this issue -- is shown in Fig.~\ref{fig:EvtHistos} with black points.

The standard MAGIC data analysis chain allows us to compensate for this via the inclusion of the energy migration effects in the spectrum reconstruction procedure. The amount of migration from each energy bin is determined from Monte Carlo simulations, updated for each MAGIC observational period. It is expressed in the form of the energy migration matrix, relating the original (``True'') energy of the gamma-ray photon to that reconstructed by the analysis (``Estimated'' energy).

This migration matrix is then used to deconvolve (or unfold) the measured event distribution and reconstruct the original spectrum of an astrophysical source. The analysis procedure allows the indicative spectral shape to be supplied, which is then used to regularise the obtained solution. The detailed procedure is described in~\citet{albert_unfolding_2007}. The MAGIC standard analysis requires several different unfolding techniques to be applied, with the result considered reliable only if all of them agree within the estimated uncertainties. These include the forward folding approach (the assumed spectral model is propagated through the MAGIC responses and its parameters are fit against the data) and three regularisation methods, further referred to as ``Bertero''~\citep{Bertero1989}, ``Schmelling''~\citep{schmelling_method_1994} and ``Tikhonov''~\citep{tikhonov1977}.

The outcome of the application of these methods is shown in Figure~\ref{fig:EvtHistos}, which summarises the measured (in terms of the estimated energy) and reconstructed (in terms of the true energy) event distributions. The true energy bins are wider than the measured energy bins, as required by the method, and show the magnitude of the spillover between energy bins.

\begin{figure}
	\centering
	\includegraphics[width=\columnwidth]{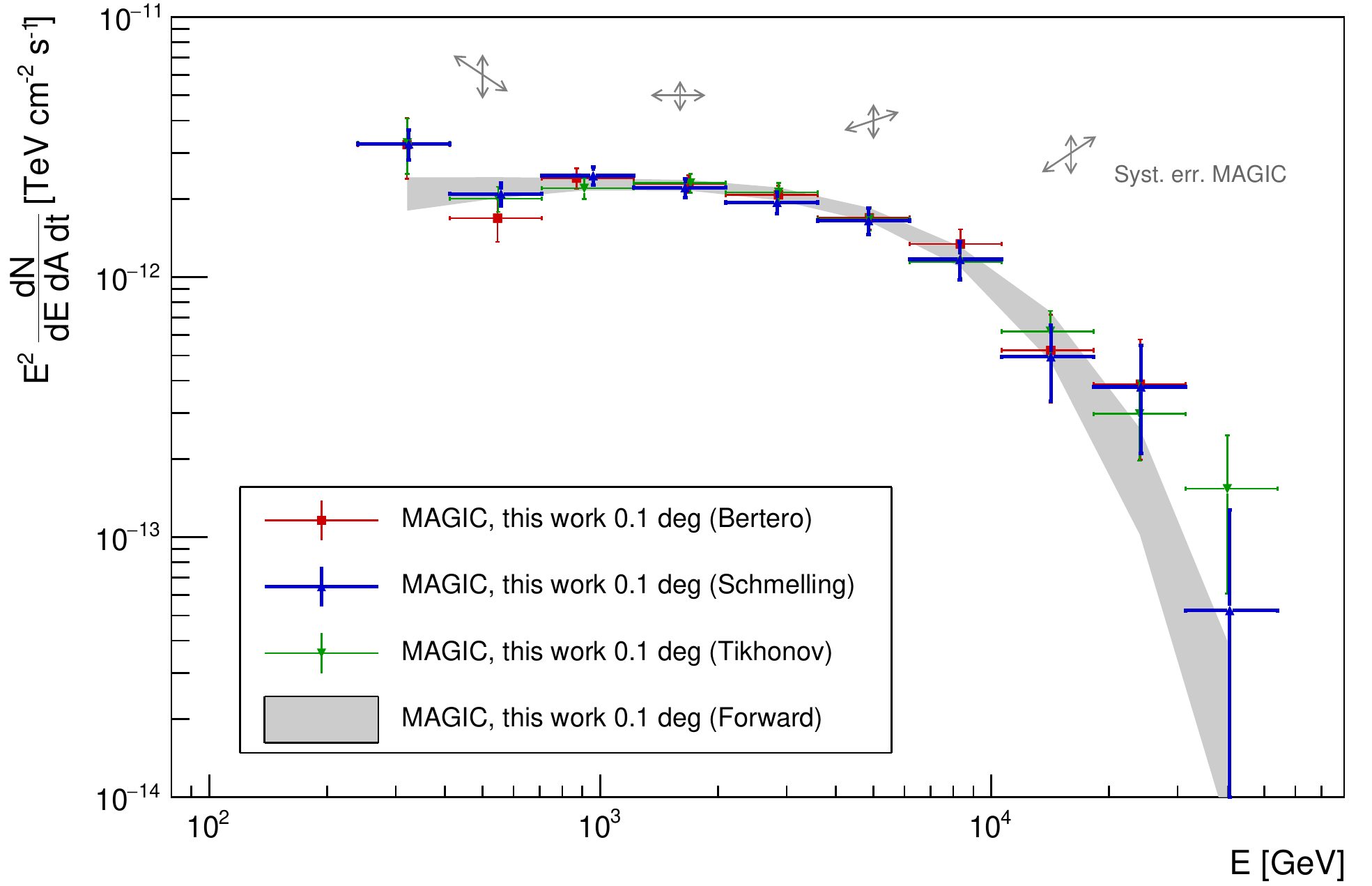}
	\caption{Unfolded SEDs of SgrA*, obtained with three different deconvolution techniques (see Appendix~\ref{sect::appendix} and Fig.~\ref{fig:EvtHistos}). The gray arrows indicate the estimated systematic uncertainty (see also Section~\ref{sect:analysis}).}
	\label{fig:UnfoldingMethods}
\end{figure}
\end{appendix}
We further used these event distributions to estimate the spectrum of the GC, corrected for the energy migration effects. The results from the unfolding techniques described above are shown in Fig.~\ref{fig:UnfoldingMethods}. All the methods yielded results that are compatible, which indicates that the determination of the true spectrum from the measured one was done reliably. For the SEDs in the main part of this manuscript (in Figs.~\ref{fig:SEDs} and~\ref{fig:SEDs_2}) we show the spectral data points obtained with the “Schmelling” technique, while, as it is commonly done in the MAGIC data analysis, the reported spectral fit results were obtained with the forward folding technique.

It is important to stress here, that the Sgr~A* observations were taken over a range of zenith distances (see Table~1 in the manuscript), where the energy threshold is changing fast, as illustrated in Section~\ref{sect:observations}. This results in a broad distribution of the detected events versus the energy, as shown in Fig.~\ref{fig:EvtHistos}. 
The lowest-energy data points in SEDs in Figs.~\ref{fig:SEDs} and~\ref{fig:UnfoldingMethods} 
are dominated by the lowest zenith angles in our Sgr~A* observations ($<60$~deg), which have the largest effective area at these energies (i.e. lowest energy threshold).

\end{document}